\documentclass[review]{elsarticle}

\usepackage{lineno,hyperref}
\usepackage{graphicx}
\usepackage{amsmath}
\usepackage{amssymb}
\usepackage{float}
\usepackage{hyperref}
\usepackage{algorithm}
\usepackage{algorithmicx}
\usepackage[noend]{algpseudocode}
\usepackage{mathabx}
\usepackage[table,xcdraw]{xcolor}
\usepackage{natbib}

\modulolinenumbers[5]

\journal{Communications in Nonlinear Science and Numerical Simulation}
\pdfoutput=1

\biboptions{authoryear}

\begin{document}

\begin{frontmatter}

\title{Precise Detection of Speech Endpoints Dynamically: A Wavelet Convolution based approach}

\author{Tanmoy Roy}
\ead{tanmoy@tanmoy.in}
\address{Electrical \& Electronic Engineering, University of Johannesburg, South Africa}

\author{Tshilidzi Marwala}
\ead{tmarwala@gmail.com}
\address{Electrical \& Electronic Engineering, University of Johannesburg, South Africa}

\author{Snehashish Chakraverty}
\ead{sne\_chak@yahoo.com}
\address{Department of Mathematics, National Institute of Technology Rourkela, India}

%
%

\begin{abstract}
Precise detection of speech endpoints is an important factor which affects the performance of the systems where speech utterances need to be extracted from the speech signal such as Automatic Speech Recognition (ASR) system. Existing endpoint detection (EPD) methods mostly uses Short-Term Energy (STE), Zero-Crossing Rate (ZCR) based approaches and their variants. But STE and ZCR based EPD algorithms often fail in the presence of Non-speech Sound Artifacts (NSAs) produced by the speakers. Algorithms based on pattern recognition and classification techniques are also proposed but require labeled data for training. A new algorithm termed as Wavelet Convolution based Speech Endpoint Detection (WCSED) is proposed  in this article to extract speech endpoints. WCSED decomposes the speech signal into high-frequency and low-frequency components using wavelet convolution and computes entropy based thresholds for the two frequency components. The low-frequency thresholds are used to extract voiced speech segments, whereas the high-frequency thresholds are used to extract the unvoiced speech segments by filtering out the NSAs. WCSED does not require any labeled data for training and can automatically extract speech segments. Experiment results show that the proposed algorithm precisely extracts speech endpoints in the presence of NSAs.
\end{abstract}

\begin{keyword}
Speech Endpoint Detection \sep Speech Recognition \sep Wavelet Convolution \sep Continuous Wavelet Transform \sep Pattern Recognition
\end{keyword}

\end{frontmatter}


\section{Introduction}\label{sec:1}
Speech endpoints are the beginning and end points of the actual speech utterance within the speech signal. Speech Recognition and its related field of research has come a long way and has matured enough. But still, precise detection of speech endpoints is an important factor affecting the recognition performance of Automatic Speech Recognition (ASR) systems. \cite{lamel} explained the importance of accurate endpoint detection in speech recognition and has shown that the speech recognition performance dramatically reduces due to an error in endpoint detection. Background noise and other sound artifacts which are not the part of the actual speech utterance exists in the speech recordings. When a recording with noise is used for analysis, the presence of those noise distorts the results. Also, the silent sections before and after the actual utterance are not required in the analysis for most of the cases, thus the requirement for precise extraction of the speech utterance by separating it from those noises and silence sections.

Digitally recorded speech can be acquired from different sources such as telephone recordings, studio recordings, conversations recorded in the natural environment. All these recordings contain various noise depending on the recording environment. Even the recordings in nearly noise-free environment contain sound artifacts produced by the speaker during the recording. Examples of such sound artifacts are mouth clicks and pops, heavy breathing and lip smacking. In this article, these sound artifacts are referred as Non-speech Sound Artifacts (NSAs). These NSAs need to be filtered out in most of the speech based applications for estimating good results because their effect is similar to noise in systems like ASR.

\begin{figure}
	\centering
	\includegraphics[width=\textwidth]{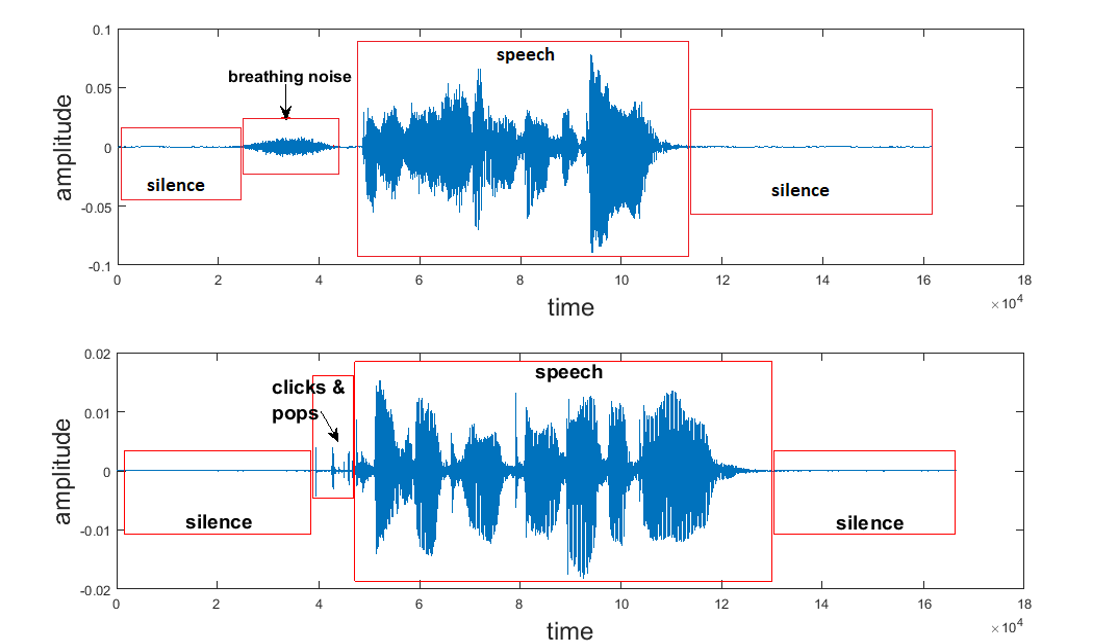}
	\caption{A speech signals containing breathing noise and mouth clicks and pops along with leading and trailing silence section.}
	\label{fig:1}       
\end{figure}

Though the quest to find a solution for End-Point Detection (EPD) problem started a long time ago in the 1970s, the search is still on because the precise solution is still not found which can cater all the difficult scenarios. Figure \ref{fig:1} shows examples of NSAs present in speech recordings such as breathing noise, mouth clicks, and pops.

\begin{figure}
	\centering
	\includegraphics[width=1.05\linewidth]{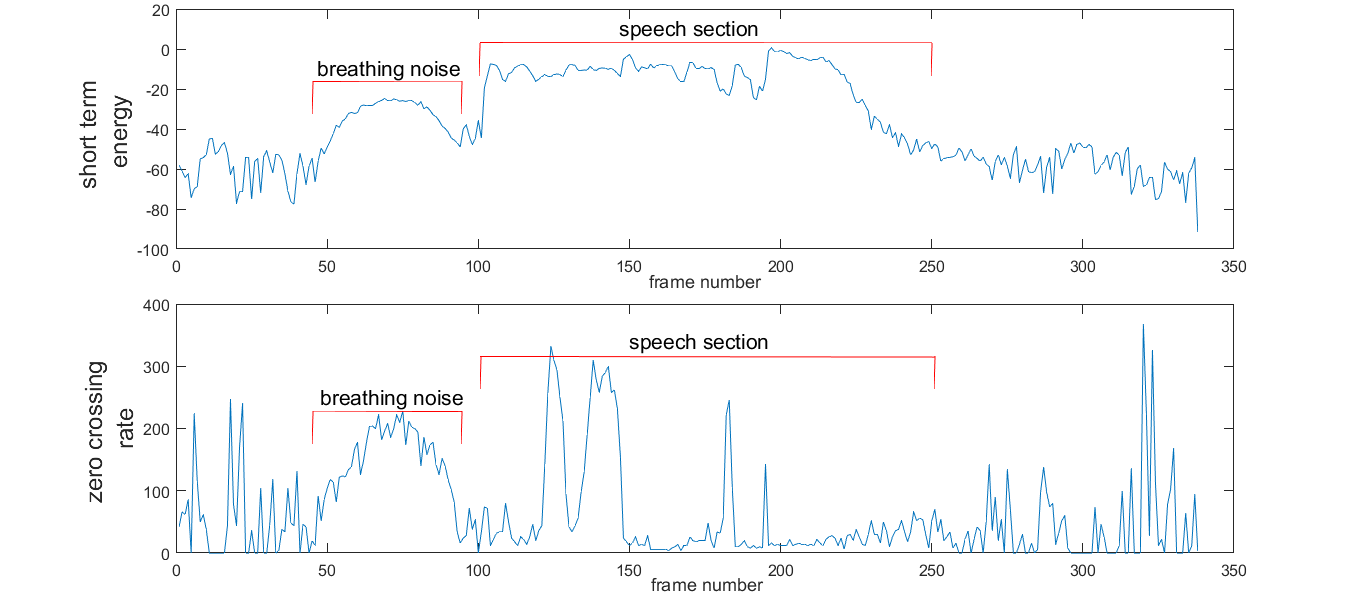}
	\caption{This figure shows how STE and ZCR plots look like in the presence of heavy breathing noise. From the plot its clear that there is not much visible distinction between the values of STE and ZCR in speech segment and noise segment.}
	\label{fig:2}       
\end{figure}

Existing EPD methods frequently use Short-Term Energy (STE) and Zero-Crossing Rate (ZCR) based methods and their variants. \cite{rabinar} proposed a simple and fast algorithm to determine endpoints based on energy and ZCR. 
\cite{savoji} also used STE and ZCR as features and their proposed algorithm uses the knowledge-based heuristics for speech classification. \cite{lamere} utilized the STE based approach with three energy thresholds, two for beginning and one for ending. Energy and ZCR based algorithms work well when there is no background noise and no NSA type noise exists in the sound recordings. Constant background noises present in speech utterances can be filtered out using a suitable noise reduction algorithm for sound. But segregating NSAs, present in the speech recordings, is a challenging task because STE and ZCR based attributes are not enough to segregate speech from NSAs. It is observed that presence of NSAs nullifies the distinction in values for STE and ZCR for speech and non-speech sections (see Fig.\ref{fig:2}). Also, \cite{lamel} have shown that energy based explicit approaches for EPD failed in the presence of NSAs. While using a heuristic approach they have classified the EPD problem into implicit, explicit and hybrid with respect to the speech recognition system. In explicit approach, EPD task is an independent module in the speech recognizer, whereas in implicit approach there is no separate stage in the recognizer for EPD. The Hybrid approach has an EPD module at the initial phase but after recognition, the initial EPD results of EPD are updated. So, when NSA type noises are present in speech utterances, STE and ZCR based approaches are not suitable for solving the EPD problem. 

Researchers have applied pattern recognition (PR) and machine learning (ML) techniques to solve EPD problem. Classification techniques such as Support Vector Machine (SVM), Hidden Markov Model (HMM), Neural Network and other suitable techniques for sequence classification are extensively used in different algorithms. \cite{atal} considered pattern recognition approach using Energy of the signal, ZCR,  Auto Correlation coefficient, First predictor coefficient, Energy of the prediction error as feature set. They also mentioned the limitations of using PR techniques. First of all, the algorithm needs to be trained for particular recording conditions. Second, manually locating voiced, unvoiced and silence for preparing training data is a tedious and time-consuming process. Hidden Markov Model (HMM) classification technique is applied by \cite{wilpon} and have shown that HMM-based EPD approach performs significantly better in the noisy environment compared to energy-based approach. \cite{qi} used the multilayer feed-forward network with hybrid features to classify voiced, unvoiced and silence from the speech and achieved 96\% classification rate. Kun et al \cite{kun} applied SVM for speech segregation in computational auditory scene analysis (CASA) problem domain and considered pitch and amplitude modulation spectrum (AMS) based features. But the presumption to work for classification techniques require properly labeled data for training and the task of labeling data is a manual or off-line process. Since manual intervention is required in the classification approach for endpoint detection, it will be difficult to automate the whole EPD process. Lamel et al \cite{lamel} also pointed that pattern classification approaches should not be readily applied in EPD owing to strong overlapping between NSAs and speech sounds. So, these are the reasons to look for techniques other than classification.

Threshold-based EPD algorithms are also proposed by some researchers. \cite{zhu} utilized the distance between autocorrelated functions and threshold as the feature set to find the endpoints. They have assumed that there exist some leading and trailing frames in the speech recording which can be considered as silence section. But this assumption might not hold for all speech databases or in real-world scenarios and that is the reason why efforts have been made to relax these assumptions.

In this article, a new algorithm is proposed as an independent module and named as WCSED (Wavelet Convolution based Speech Endpoint Detection). The WCSED algorithm is a deviation from the energy and ZCR based approaches. It is formulated by utilizing the simple fact that NSAs are high-frequency sound, and used the concepts of wavelet convolution and entropy as a building block. First, the input speech signal is decomposed into high-frequency (HF) and low-frequency (LF) components using wavelet convolution method. It is observed (Fig \ref{fig:4}) that the NSAs are much prominent in the HF components than in the LF components. Also, the voiced sections of a speech utterance are low-frequency sounds whereas unvoiced sections are high-frequency sounds. Thus it can be stated that the HF components represent both the unvoiced speech and the NSAs, and the LF components represent the voiced speech. Two sets of thresholds are computed based on the entropy values for both the HF and LF components. The speech signal is broken down into manageable frames to calculate the entropy of the decomposed components. The LF thresholds extract the voiced speech segment whereas the HF thresholds are used to segregate the unvoiced speech segments from the NSAs. Results show that WCSED precisely extracts speech segments in the presence of NSAs. Moreover, the proposed algorithm works with unlabeled data as there is no training involved. Which contributes to the easy automation of the EPD process by the proposed algorithm. Also, in WCSED, threshold computation do not assume that there exists a fixed number of leading and trailing frames, which further improves the flexibility of the algorithm as far as the use of dataset is concerned.

This article is organized into following sections. Section \ref{sec:2} describes the problem in hand. Section \ref{sec:3} dedicated to describing the proposed solution in detail and relevant concepts are also discussed. Section \ref{sec:4} briefly describes the dataset we used. In Section \ref{sec:5} results of the algorithm and observations are elaborated. And finally Section \ref{sec:6} concludes this article and mentions possible directions which can be explored to extend or utilize this work.

%

\section{The Problem}\label{sec:2}
In this section, the problem of speech endpoint detection is elaborately described.
\subsection{Difficulties in endpoint detection}\label{subsec:2.1}
Continuous speech signals are recorded, digitized and stored as discrete time signals which are mostly used for speech-based applications such as ASR, Speech Emotion Recognition (SER) etc.

Apart from speech segment, speech recordings contain two more segments, the silence section at the beginning and at the end of the recordings and the noise section (see Fig.\ref{fig:1}). Speech databases from different projects are recorded with a different degree of background noise. Here we are considering speech database which is recorded in a quiet environment with very little or no continuous background noise. Although there is negligible background noise, there are some unwanted sound artifacts got generated during the course of recording by the speakers such as lip smacking, heavy breathing, mouth clicks, and pops. Fig.\ref{fig:1} shows the presence of NSAs in speech recording.

The problem here is to separate speech utterances from silence and noise segments. Silence can be usually separated by applying algorithms based on STE and ZCR when there is negligible continuous background noise and no NSAs exists in the recordings. 
But STE based approaches fail to segregate the energy level of speech and noise when noise exists in recordings. Moreover, noise and speech segments of a recording don't contain any standard characteristics which can distinguish them. Also, human speech contains two types of sound, Voiced sounds such as vowels (a,e,i,o,u) 
and unvoiced sounds such as k and p. 
The characteristics of unvoiced sounds are very similar to noise and that needs to be taken care of while filtering out the noise.
So the problem here has three folds 
\begin{itemize}
	\item segregate speech from trailing and leading silence
	\item consider the presence of noise
	\item need to be careful about not to consider unvoiced speech sounds as noise.
\end{itemize}

\subsection{Problem Statement}\label{subsec:2.2}

We are considering discrete-time speech signals as input to our system. A discrete time signal X can be mathematically represented as a sequence of numbers as follows:

\begin{equation}\label{eq:1}
\begin{aligned}
X = \{x[n]\}, \hspace{2mm} where \hspace{1mm} & x[n] = \{x_1,x_2,...,x_n\}, \\
& -\infty < n < \infty, \\
& (x_1,x_2,...,x_n) \in \mathbb{R} \\
\end{aligned}
\end{equation}

here n is an integer and $x[n]$ is the sequence usually generated by taking a periodic sample from an analog signal.

\begin{equation}\label{eq:2}
\begin{aligned}
x[n] = \{idle[k],speech[m],noise[l] \},
where \hspace{2mm} & n = k + m + l
\end{aligned}
\end{equation}

This sequence $x[n]$ comprises of three sections (eq \ref{eq:2}), the idle section $idle[k]$, the noise section $noise[l]$ and the speech section $speech[m]$ where $n = k + l + m$. These sections are not distinguishable by mere evaluation of the values in these sequences because no predefined ranges or thresholds of values exists.

The task here is to extract only the $speech[m]$ section from $x[n]$. It is assumed here that $speech[m]$ contains a continuous sequence extracted from $x[n]$. But the $noise[l]$ and $idle[k]$ sections can contain combination of multiple sequence fragments from $x[n]$. So, the sequence of $x[n]$ contained in $speech[m]$ cannot be found in either $noise[l]$ or in $idle[k]$.

So, the objective here is to look for pattern in $x[n]$, that can distinguish $speech[m]$ from $noise[l]$ and $idle[k]$ and finally extract the $speech[m]$ from $x[n]$.


\section{The Proposed Solution}\label{sec:3}
A solution based on wavelet convolution to the problem stated in section \ref{subsec:2.2} is proposed here. The pattern has been found in the speech signals that demarcate speech utterances from a non-speech section of the recording. The concept of entropy is applied to get an approximation of information content in wavelet convolution coefficients. In the following subsections, these concepts are discussed before formulating the actual solution.

\subsection{Convolution}\label{subsec:3.1}
Convolution is an important operation in signal and image processing domain. It is a concept extensively used in linear algebra. Convolution is one of the cornerstones of wavelet transform concept and continuous wavelet transform is applied to solve the endpoint detection problem. In this section concept of continuous convolution is briefly discussed.

Convolution operates with two functions, one is $input$ and another is $kernel$, and produces a third function. First, the $kernel$ is flipped (rotation by 180 ) about its origin and slided past the $input$ to compute the sum of products at each displacement. Let there be an input function $f$ and kernel function $g$. Then the convolution between $f$ and $g$, denoted by $h$, is defined as follows:



\begin{equation}\label{eq:3}
h(i) = (f \convolution g)(i) = \int_{-\infty}^{\infty}f(i-j)g(j)dj
\end{equation}

where the minus sign accounts for the flipping of the kernel function $g$, i is the required displacement and j is a dummy variable that is integrated out \cite{gonzalez}. 

\subsection{Wavelets}\label{subsec:3.2}
The Concept of Wavelet decomposition is the key to solving the speech endpoint detection problem in this algorithm. This section described important and relevant areas of the Wavelet concept in as much detail required for this work.
\subsubsection{Why Wavelets?}\label{subsubsec:3.2.1}
Signals carry overwhelming amounts of data which needs to be extracted as information. But often the difficulties involved in the task of extracting relevant information from those data becomes a hurdle for the field of study to which those signals belong. Sparse representation of signals is an efficient way to look for relevant information and patterns in signals. Sparse representation is achieved through decomposing signals over oscillatory waveforms using Fourier or wavelet bases. Speech signals too carry different types of data that need to be extracted as information for better results in various research areas and applications that uses speech signals.

Non-stationary signals are the signals whose frequencies and other statistical properties varies over time. Fourier Transform (FT) is not suitable for analyzing non-stationary signals. Short Time Fourier Transform (STFT) was introduced to overcome this shortcoming of FT. But during STFT process while transforming time domain signal into frequency domain vital time information is lost. This phenomenon of losing time information can be explained by Heisenberg's Uncertainty Principle [see \cite{mallat}].

Wavelet analysis is best suited in this scenario where we have to analyze the non-stationary signal to look for a change in frequency components over time. Speech is a non-stationary signal. For this reason, wavelet decomposition is applied here to find relevant frequency components in speech signals. Wavelets define a sparse representation of well-localized piecewise regular signal through the coefficient amplitudes and few coefficients are required to represent that transient structure. That sparse representation may include transients and singularities. This why wavelet analysis is important in speech processing.

\subsubsection{Wavelet Analysis}\label{subsubsec:3.2.2}

This section describes the method of wavelet analysis. Consider a finite energy signal $x(t)$ where the energy of $x$ is defined by its squared norm and is expressed as 

\begin{equation*} 
\Vert x(t) \Vert ^2 = \int_{-\infty}^{+\infty} \vert (x(t)) \vert ^2 dt < +\infty
\end{equation*}

So, the space on which the $\Vert x(t) \Vert ^2$ norm is defined has to be square integrable because the integral $ \int_{+\infty}^{-\infty} \vert (x(t)) \vert ^2 dt $ must exists. That space is denoted as $\mathbb{L}^2(\mathbb{R})$ is a Hilbert space and is the vector space of the finite energy functions and thus $x(t) \in \mathbb{L}^2(\mathbb{R})$.

The objective here is to decompose the signal $x$ into a linear combination of a set of functions which belongs to $\mathbb{L}^2(\mathbb{R})$. Let us consider a function $ \psi(x) \in\mathbb{L}^2(\mathbb{R})$ whose dilation and translation forms a set of functions in $\mathbb{L}^2(\mathbb{R})$ space

\begin{equation*}
\psi_{\tau,s}(t) = \frac{1}{\sqrt{s}} \psi \left(\frac{t - \tau}{s}\right), \hspace{1mm} where \hspace{1mm} \tau \in \mathbb{R}, s \in \mathbb{R}^+ \hspace{1mm} and \hspace{1mm} s\neq0
\end{equation*}

$\tau$ and $s$ are the translation and scaling (dilation) parameters respectively and $s$ cannot be negative since negative scaling is undefined. Normalization by $\frac{1}{\sqrt{s}} $ ensures that $\Vert \psi_{\tau,s}(t) \Vert$ is independent of $s$. The family of functions $\psi_{\tau,s}$ is called \textit{wavelets} and $\psi$ is called the \textit{mother wavelet}.

So, now the signal $x$ can be represented as wavelet inner-product coefficients

\begin{equation}\label{eq:4}
\langle x, \psi_{\tau,s} \rangle = \int_{-\infty}^{\infty} x(t) \psi_{\tau,s} (t) dt 
\end{equation}

here both $x$ and $\psi$ are considered as real-valued signals. When $\psi$ is a complex wavelet, the right hand side of equation \ref{eq:4} will have complex conjugate of $\psi$ as $\psi^*_{\tau,s} (t)$. The \textit{mother wavelet}, also referred to as the \textit{wavelet function} or the \textit{kernel function}, has zero average, meaning $\int_{-\infty}^{\infty} \psi(t) dt = 0$. Apart from satisfying zero average condition \textit{wavelet functions} has to satisfy two more mathematical criteria. First one is that the wavelet function must have finite energy: $E = \int_{-\infty}^{\infty} |\psi(t)|^2 dt < \infty$, which ensures that $\psi$ is square integrable and the inner product in eq \ref{eq:4} exist. And the second one is called the admissibility condition which eventually boils down to the condition of zero average, stated earlier, which ensure that $x$ can be reconstructed again after decomposition. The \textit{wavelet function} need to be selected carefully based on the type of analysis to be performed on the input signal because that will help to identify regularities and singularities. The choice of the \textit{mother wavelet} to be used in continuous wavelet transform is restricted only to the conditions of finite energy and admissibility \cite{daubechies}. \textit{Wavelet function} can be either orthogonal or nonorthogonal and only the orthogonal functions form \textit{wavelet basis}. That is why the orthogonal wavelets give compact representation of the signal and are useful for signal processing. On the other hand nonorthogonal wavelets produce wavelet spectrum which is highly redundant at high scales and are more useful for time series analysis (\cite{torrence}).

Here continuous wavelet transform (CWT) is used for the analysis, so we will concentrate on CWT. But before going into details of CWT here are two reasons behind selecting CWT over Discrete Wavelet Transform (DWT) for this solution. \cite{mallat} mentioned, discrete sequence of $\tau$ is complex to describe and amplitudes of wavelet coefficients are difficult to interpret since the regularity of a discrete sequence is not well defined. Moreover, the purpose of the CWT is to extract information from signal whereas DWT is good at reconstructing the signal. Here information needs to be extracted from speech signals and thus CWT is chosen. The scaling parameter $s$ in CWT can vary continuously over $\mathbb{R}$ and can take any value, whereas values $s$ are restricted in DWT. So, signal analysis at arbitrary scale (or frequency) is possible in CWT and not in DWT, which is an important criteria for the current problem.

Now, CWT of $x(t)$ with respect to wavelet function $\psi(t)$ at scale $s$ and position $\tau$ is the projection of $x$ on $\psi$ and is defined as inner product coefficients in eq \ref{eq:4}:

\begin{equation*}
C(\tau,s;x(t),\psi(t)) = \langle x, \psi_{\tau,s} \rangle =  \int_{-\infty}^{\infty} x(t) \psi_{\tau,s}(t) dt
\end{equation*}

which measures the variation of $x$ in the neighborhood of $\tau$ proportional to $s$.
\cite{calderon} has shown that CWT can be defined as a convolution operation.

\begin{equation}\label{eq:5}
C(\tau,s;x(t),\psi(t)) = \int_{-\infty}^{\infty} x(t) \psi_{\tau,s}(t) dt = x \convolution \bar{\psi}(\tau)
\end{equation}

where

\begin{equation*}
\bar{\psi}(\tau) = \frac{1}{\sqrt{s}} \psi \left(\frac{-t}{s}\right)
\end{equation*}

So, CWT extracts information by convolution and not exactly decomposes the signal into sub-signals. Since CWT uses non-orthogonal wavelets, reconstruction frame is less important and problematic as well because the inverse wavelet transform for CWT is still not well defined. This wavelet convolution operation is the foundation of the proposed solution.

CWT must be discretized to be implemented in a computer. That is what is done here by selecting a discrete set of relevant scales for analysis rather than continuous scale. The shifting (translation) has to be done continuously over for all the points of the signal to be analyzed through convolution operation as defined in Eq \ref{eq:3}.

\subsection{Entropy}\label{subsec:3.3}
Entropy was introduced in physics as a thermodynamic state variable. It provides an appropriate measure of randomness or disorganization in a system and increases along with the randomness of the system. Statistically its defined as (see. \cite{kullback}):

\begin{equation}\label{eq:6}
E(X) = \sum_{i=1}^{N} p(x_i)log_{10}p(x_i),
\end{equation}

where $X = \{x_1,x_2,...,x_N\}$ is a set of random phenomena, and $p(x_i)$ is the probability of a random phenomenon $x_i$.

During this work, its observed that entropy of amplitude values of a signal continues to be significantly high and stable when there is descent disturbance in the system. This is a useful observation to keep track of voice activity in a signal recording and separate voice from silence. So, from the current problem perspective described in section \ref{subsec:2.2} we can write

\begin{equation}\label{eq:7}
E(speech[m]) \gg E(idle[m])
\end{equation}

In the proposed algorithm the concept of entropy is a key component in separating speech section from silence.

\subsection{Concept of Frame}\label{subsec:3.4}
Human speech generation apparatus that is tongue, lip and the other parts of our vocal system involved in producing sound needs approximately 25-30 milliseconds gap between two uttered words because it needs that time to prepare the system to produce next sound. So, if it is required to break the signal into smaller frames the size should be chosen within that range. Frames are needed for this algorithm and its fixed at \textbf{20ms} and is termed as \textit{frame length}. Also, the concept of \textit{frame shift} is used to define the actual shift of data points in the signal, which is fixed at \textbf{10ms}. Combination of \textit{frame length} and \textit{frame shift} is used to avoid the effect of the abrupt split of waves during frame splits, to some extent.

\subsection{Formulation of the Solution}\label{subsec:3.5}
\begin{figure}
	\centering
	\includegraphics[width=\textwidth]{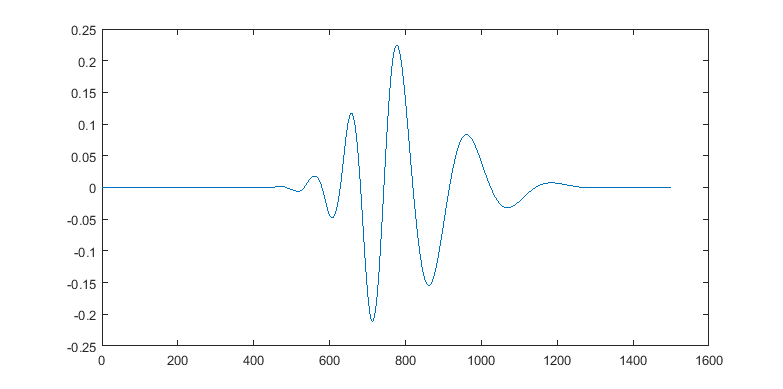}
	\caption{Figure shows $DB_8$ wavelet shape at scale 100}
	\label{fig:3}
\end{figure}

The first step to apply wavelet decomposition method for analysing asignal is to select a suitable \textit{mother wavelet}.
Here Daubechies wavelet have been selected for this algorithm, specifically $DB_8$. Daubechies wavelets are one of the popular wavelets among researchers for speech processing (\cite{tan},\cite{campo}). Shape of a $DB_8$ signal is shown in Fig.\ref{fig:3}. Since continuous wavelet transform is considered here, the scaling and translation parameters $s$, $\tau$ can vary continuously over $\mathbb{R}$ \cite{daubechies}. So, from continuous scales, arbitrary set of scales is selected to cover the possible frequency range of the human speech recording signals. Here an orthogonal wavelet function $DB_8$ is convolved over the discrete input signal to get the coefficient values at different scales (frequencies). Orthogonality of $DB_8$ helps to remove the redundancy of wavelet coefficient.

\begin{figure}
	\centering
	\includegraphics[width=\textwidth]{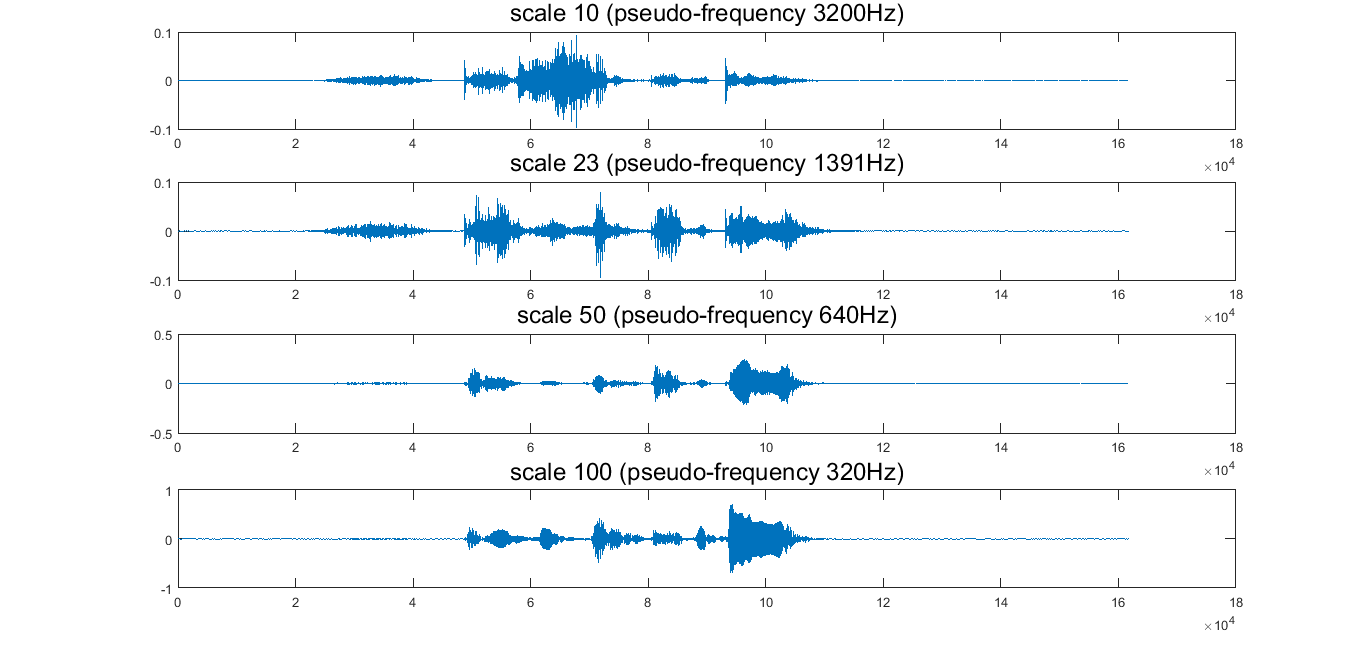}
	\caption{Coefficient Amplitudes at different scales for a speech utterance with breathing noise. Scale 10 highlights 3200Hz frequency components where breathing noise is very prominent. Scale 23 highlights 1391Hz frequency components where noise is most prominent compared to a speech utterance. Scale 50 highlights 640Hz frequency components where the weak presence of noise can be seen. And finally, Scale 100 highlights 320Hz frequency components where the noise section is very weak compared to speech section.}
	\label{fig:4}       
\end{figure}

Objective here, as described in Section \ref{subsec:2.2}, is to find pattern in discrete sequence $x[n]$ (eq \ref{eq:1}) to segregate speech segment from rest of the sequence. Wavelet convolution operation is applied to analyze the sequence $x[n]$ and search for relevant patterns. Its observed during the experiments that presence of NSAs are prominent in coefficient amplitude plot when wavelet scale is small (high-frequency) (Fig.\ref{fig:4}). It is equivalent to the fact that NSAs has similarities with high-frequency wavelets since low scale value implies high-frequency. But as we go on analyzing the coefficients in higher scales (low-frequencies) we found that those NSAs are almost non-existent in the plot (Fig.\ref{fig:4}). The phenomenon is well supported by the fact that NSAs are usually high-frequency sounds 
and thus produces high coefficient values in convolution with low scale (high-frequency) wavelets. This observed phenomenon is the backbone of this approach to solving the problem of speech endpoint detection.

A set of scales has been selected based on the range of frequency we need to cover. The frequency components of the human speech signal are mostly covered within the range between 250Hz and 6000Hz (\cite{shen}). But its observed that NSAs are prominent around 3000Hz and around 300Hz the presence of noise is very weak, so here we will consider the upper limit as 3000Hz and lower limit as 300Hz. To accommodate that frequency range using $DB_8$ mother wavelet two sets of scales are selected: 
\begin{itemize}
	\item $scale_{hf}$ includes set of high frequency range (low scale values)
	\item $scale_{lf}$ includes set of low frequency range (high scale values)
\end{itemize}
At low scale, wavelet coefficient values are much smaller compared to coefficient values at high scale. This is the reason why more number of scales are selected for $scale_{hf}$ than $scale_{lf}$.

Its assumed here that there exists a gap of few milliseconds between the NSAs and the speech utterances. It is very unlikely that the speaker can produce some NSAs exactly before and after the actual utterance without any time gap. For example, the noise of breathing out cannot come out while speaking because the voice is already coming out with exhalation, and if at all breathing noise comes out while speaking it would distort the speech utterance. Similarly, mouth pop and click sounds cannot be produced by the speaker while uttering a speech because that will interrupt the utterance.

Now wavelet transform of the discrete sequence $x[n]$ (\ref{eq:1}) is performed, which is defined as convolution of $x$ with a scaled and translated version of $\psi$ the mother wavelet ($DB_8$) (\cite{torrence}) to generate set of coefficients as described in eq \ref{eq:5}.

\begin{equation}\label{eq:8}
coefs = x \convolution \psi
\end{equation}	

Coefficient sets are needed to be combined together to get two vectors that can be used for further processing. To achieve that sum or average strategy has been applied depending on the loudness of the actual signal X. When loudness is higher than a specific threshold value, the coefficient values are averaged otherwise they are summed.

After the coefficients are combined into two vectors namely $coef_{hf}$ and $coef_{lf}$, the entropy is computed for both the vectors. The coefficient vectors are broken down into frames and then entropy is computed using the formula defined in eq \ref{eq:6}. These entropy vectors are special in a sense that they represent high-frequency entropy (say $ce_{h}$) and low-frequency entropy (say $ce_{l}$) of the wavelet coefficients.

The entropy vectors $ce_{h}$ and $ce_{l}$ are further used to calculate two sets of thresholds one for high-frequency and the other for low-frequency. Low-frequency thresholds are used to identify locations with presence of speech utterance because low-frequency components are distinctly separate from $idle[k]$ and $noise[l]$ sections. Then high-frequency thresholds are used to stretch those identified speech utterance zones with proper voiced and unvoiced trails at the beginning and end of speech utterance.


\subsection{The Algorithm}\label{subsec:3.6}
The proposed algorithm WCSED is designed to work independently. Systems require extracting speech segment from speech signals can incorporate this as a separate module.
The steps of the proposed algorithm are listed in Algo \ref{algo:1} section. Here the pseudo code is provided in the listing and the functions, in brief, are mentioned to maintain the readability of the algorithm.

WCSED algorithm consists of one main module and three submodules. The main module called WCSED which accepts discrete time speech signal as input and returns the extracted speech segment. The "WaveConv" module is responsible for computing the CWT on the input signal and returns the coefficients. The "GetEntropyVector" module computes entropy by breaking down the input sequence into segments and returns a vector. And finally, the points towards the edges of the end-points are selected by considering the threshold values provided.

Assumptions for this WCSED algorithm are kept at the minimum to maintain generality. Thresholding concept was applied but the assumptions of leading and trailing silence similar to Zhu et al \cite{zhu} is relaxed because that would restrict the scope of this algorithm to specific datasets. Thresholds are dynamically calculated.

\begin{figure}
	\centering
	\includegraphics[width=0.8\textwidth,height=3in]{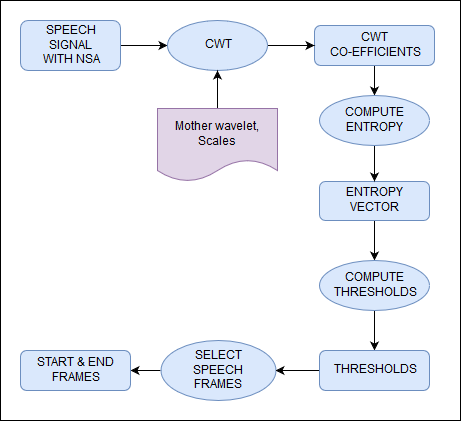}
	\caption{Block diagram of the WCSED algorithm}
	\label{fig:7}
\end{figure}

\begin{algorithm}
	\caption{WCSED algorithm}
	\label{algo:1}
	\textbf{Input:} Discrete-time signal S(n), where n is the length of the signal and Sampling Rate
	\textbf{Output:} Extracted Speech Segment $S_{extr}$(k), where $k<=n$
	\begin{algorithmic}[1]
		\Function{WCSED}{$S(n),FS$}\Comment{S=discrete time signal and FS=sampling rate}
		\State $FL \gets Frame Length$ 
		\State $FSH \gets Frame Shift$
		\State $MW \gets "Daubechies"$ \Comment mother wavelet
		\State $SC(m) \gets [HighFrequencyScales,LowFrequencyScales]$ \Comment m number of scales
		\State $COEF_{m \times n} \gets WaveConv(S(n),SC(m),MW)$ \Comment coefficients
		\State $CE \gets GetEntropyVector(COEF_{m \times n},FL,FSH)$
		\State $th_{u},th_{l} \gets $ compute upper and lower thresholds  
		\State $sec_{al} \gets CE \ge th_{l}$
		\State $pos_{s} \gets IncludeEdges(CE,sec_{al}[start],back,th_{u})$
		\State $pos_{e} \gets IncludeEdges(CE,sec_{al}[end],front,th_{l})$
		\State $S_{extr}(k) = S[pos_{s},pos_{e}]$
		\State \textbf{return} $S_{extr}(k)$\Comment{The extracted speech}
		\EndFunction
	\end{algorithmic}
	
	\begin{algorithmic}[1]
		\Function{WaveConv}{$S,SC,MW$} \Comment{signal,scales,mother wavelet}
		\State $CD_{m \times n} \gets $ output matrix
		\For{($m=1;m<=lenght(SC);m++$)} \Comment{iterate through all the scales}
		\State $f \gets $ get the reference wavelet
		\State $CF_{1 \times n} \gets S\convolution f$ \Comment{convolution gives the coefficients}
		\State $CD_{m:} \gets diff(CF)$ \Comment{take approximate derivative}
		\EndFor
		\State \textbf{return} $cd$ \Comment{derivative of coefficients}
		\EndFunction
	\end{algorithmic}
	
	\begin{algorithmic}[1]
		\Function{GetEntropyVector}{$i,fl,fs$} \Comment{input sequence, frame len, frame shift}
		\State $len \gets length(i)$
		\State $sp \gets 1$
		\State $ep \gets sp + fl - 1$
		\While{$ep \leq len$}
		\State $entropy_v \gets Entropy(i[sp,ep])$\Comment calculate entropy
		\State $sp \gets sp + fs$
		\State $ep \gets sp + fl - 1$
		\EndWhile\
		\State \textbf{return} $entropy_v$\Comment{The entropy vector}
		\EndFunction
	\end{algorithmic}
	
\end{algorithm}

\section{Dataset}\label{sec:4}
This study is based on Ryerson Audio-Visual Database of Emotional Speech and Song (RAVDESS) \cite{ravdess} dataset. This dataset was primarily created in view of research areas related to  Emotion Recognition in Speech and Song. Only the speech recordings are used for this current work. While working on Speech Emotion Recognition its observed that the recordings contain different sound artifacts generated by the speakers such as heavy breathing, mouth clicks and pops, lip-smacking. These sound artifacts are making endpoint detection task difficult and the need for a robust endpoint detection algorithm was felt. There are total 24 speakers of which 12 male and 12 female. The speakers utter two statements "kids are talking by the door" and "dogs are sitting by the door". The utterances are varying over different emotions and intensities. 

\section{Results and Observations}\label{sec:5}

\begin{figure}
	\centering
	\includegraphics[width=\linewidth]{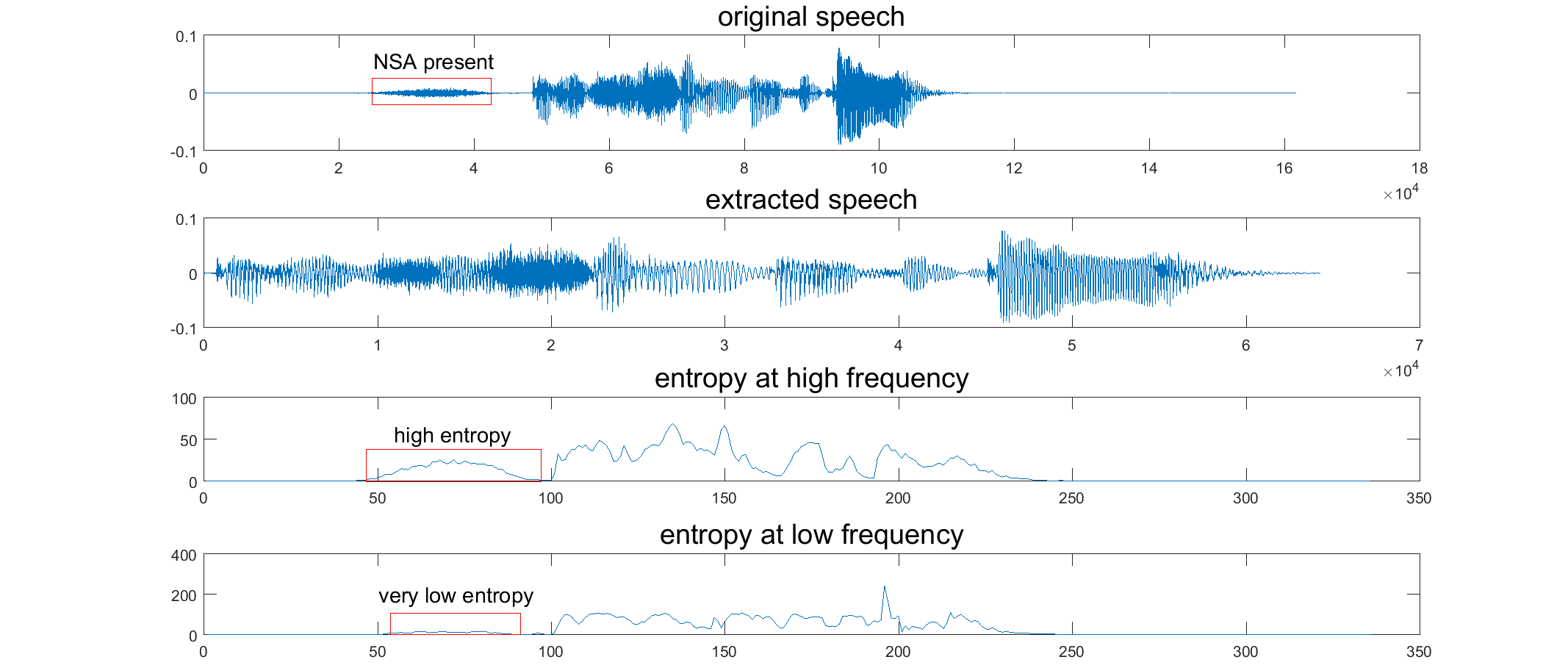}
	\caption{The Figure shows extracted speech along with corresponding entropy. The breathing noise NSA is precisely discarded.}
	\label{fig:5}
\end{figure}

\begin{figure}
	\centering
	\includegraphics[width=\textwidth]{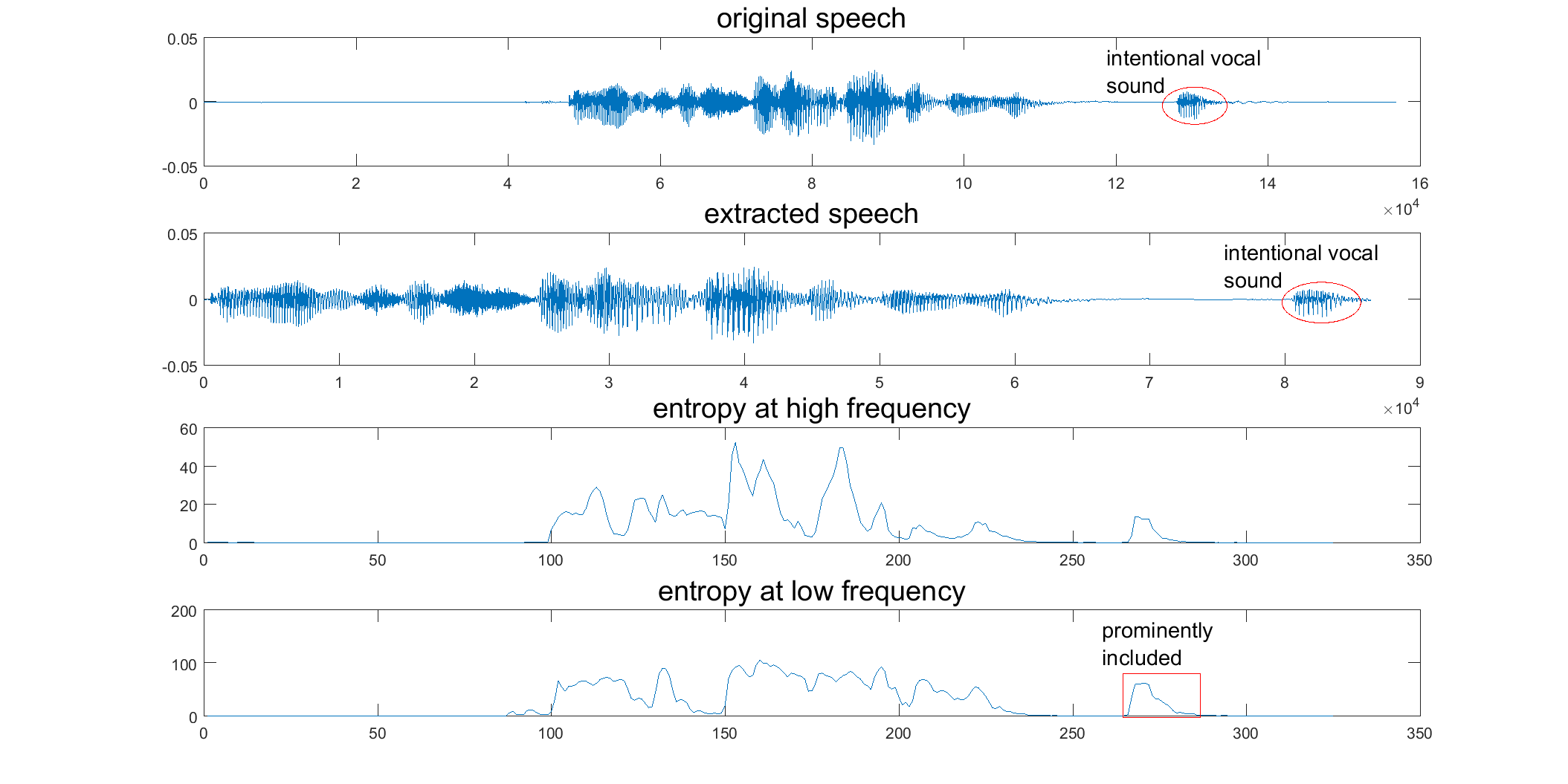}
	\caption{The Figure shows extracted speech along with corresponding entropy. Speaker's intentional voice sound is meaningfully included in the extracted speech.}
	\label{fig:6}
\end{figure}
The primary objective of WCSED algorithm is to automate the process of extracting the speech segments precisely in the presence of NSAs and it has shown promising results. It has successfully extracted the speech segments from almost all the recordings. In very few cases the significant amount of speech could not be extracted but the algorithm did not fail completely in those rare cases. The speech recordings containing NSAs are efficiently processed by separating those unwanted artifacts from actual speech.

Some speakers pause for some few milliseconds between the words. Those pauses should be included as a part of speech segment since pauses can add quality to the speech recording while extracting say emotional quotient and the algorithm did it well in those cases too.

Fig.\ref{fig:5} and Fig.\ref{fig:6} show the end result of the algorithm depicting the extracted segment along with corresponding entropy values.

\begin{table}[H]
	\centering
	\caption{Test Results}
	\label{tab:1}
	\begin{tabular}{|c r r|}
		\hline
		\rowcolor[HTML]{BBDAFF} Speaker &Average \% of &Average \% of \\
		\rowcolor[HTML]{BBDAFF} Gender &Startpoint Deviation &Endpoint Deviation \\
		FEMALE &1.027 &2.259 \\ 
		MALE &0.576 &2.847 \\ 
		\rowcolor[HTML]{BBDAFF} Average \% of deviation & \textbf{0.777} & \textbf{2.584} \\
		\hline
	\end{tabular}
\end{table}

The experiment results are summarized in Table \ref{tab:1}, where the deviations are depicted. More than 20\% of the total number of speech recordings are selected as sample for cross verifying with the results received by applying the WCSED algorithm. Those samples are manually checked for possible start-frames and end-frames of the speech segments in the recordings. Since WCSED algorithm extracts speech segment based on frames, the selected samples are also processed based on start and end frames. After manually extracting the frames of the samples it is checked that how the start and end frames are deviating from the frames reported by WCSED algorithm of the corresponding speech recordings. 

Simulations are executed 10 times on the selected sample to check whether there is any discrepancy in different simulations. But its observed that in every simulation the algorithm has produced exactly same results. The cross verification of results is measured in few stages. First, beginning and end frames are calculated for the selected samples manually, let us refer them as manual-frames. Those manual-frames are then compared with corresponding frames reported by the WCSED algorithm, let us refer them as algorithm-frames. Then absolute deviation between manual-frames and algorithm-frames are computed. Considering manual-frames as a baseline, the length of the extracted speech is calculated and then the percentage of deviation in frames, compared to the frame length of the extracted speech, is calculated. This percentage deviation is depicted in Table \ref{tab:1}.

Analyzing the deviations its observed that overall start-frame deviation is 0.777\% (means approximately 99.3\% accurate), while end-frame deviation is 2.585\% (means approximately 97.5\% accurate). Thus, the algorithm extracts the start frames more accurately than the end frames. This accuracy gap is due to the fact that different speakers end their utterance with different styles and varying pause or silence between spoken words. So, the overall accuracy of the WCSED algorithm to detect start-frame is 99.3\% (approx) and end-frame is 97.5\%(approx). 

It is observed during testing that the deviations are different for female and male speakers. Factor contributed to this phenomenon is possibly the loudness variation in female and male speakers, male voices in this recordings are usually louder and more prominent than female voices.

Finally, the time complexity of WCSED algorithm is directly proportional to the length of the input signal. When input signal length increases, the algorithm will take more time to extract the speech utterance from the input signal.

\section{Conclusion}\label{sec:6}
The proposed WCSED algorithm tried to address four issues of speech end-point detection problem. First, automating the process of EPD. Second, discarding the NSAs and extracting start and end points properly. Thirdly, relaxing assumptions which could hinder this algorithm to work properly across different speech databases and in real-world applications. Finally and most importantly extract the end-points accurately. The results discussed in section \ref{sec:5} are promising and WCSED is able to address the aforementioned issues.

This algorithm can be further applied in different speech signal based systems where utterances need to be extracted from speech signals in the presence of different NSAs. For example, this algorithm can be applied in the preprocessing stage of an ASR or an SER system.

Wavelet convolution (CWT) based approach to find relevant patterns in a discrete time signal can be applied to solve similar problems in speech recognition domain and other domains where patterns need to be identified from signals. CWT can be used to enhance the feature set of various classification problems.

Result \ref{sec:5} section mentioned that level of loudness of speaker's utterances could be an important factor to improve the end-point selection results. Further investigation and action in that direction could yield more accuracy from this WCSED algorithm.


\bibliography{references}

\end{document}